\documentstyle[12pt,aas2pp4]{article}
\def\msun{M$_\odot$}

\begin{document}

\title{Gravitational Interactions in Poor Galaxy Groups}
\author{David S. Davis\altaffilmark{1} and William C. Keel\altaffilmark{2}}
\affil{Department of Physics and Astronomy,
The University of Alabama, 
206 Gallalee Hall, 
Tuscaloosa AL 35487}

\author{John S. Mulchaey}
\affil{The Observatories of the Carnegie Institution of Washington,
813 Santa Barbara St., 
Pasadena CA 91101-1292}

\and

\author{Patricia A. Henning}
\affil{Department of Physics and Astronomy,
The University of New Mexico,
800 Yale Blvd., NE,
Albuquerque, NM 87131}

\slugcomment{To be submitted to the Astronomical Journal}
\altaffiltext{1}{Present address: MIT Center for Space Research, Building
37-662B, Cambridge, MA 02139-4307}
\altaffiltext{2}{Visiting astronomer, Kitt Peak National Observatory, National
Optical Astronomy Observatories, operated by AURA, Inc., under cooperative
agreement with the National Science Foundation.}

\begin{abstract}

We report the results of the spatial analysis of deep 
{\it ROSAT} HRI observations, optical imaging and spectroscopy, and 
high--resolution VLA H I and continuum imaging 
of NGC~1961 and NGC~2276. These spirals were selected as showing some previous
evidence for interaction with a surrounding (hot) diffuse medium.

Our results favor most aspects of these galaxies as being shaped by
gravitational interactions with companions, rather than the 
asymmetric pressure from motion through an external medium.
The old stars follow the asymmetric structures of young stars
and ionized gas, which suggests a tidal origin for the lopsided
appearance of these galaxies. In NGC 2276, the H I and star-forming regions
are strongly concentrated along the western edge of the disk. In this
case, the ROSAT HRI detects the brightest star-forming regions as
well as the diffuse disk emission, the most distant galaxy with such
a detection. An asymmetric ionization gradient in the H II regions
suggests radial movement of gas, which might have occurred in either
tidal or wind scenarios. The X-ray emission from NGC~1961 is dominated
by a point source near the nucleus of the galaxy but extended emission 
is seen out to a radius of $\sim$ 0\farcm 8. 

Previous studies of 
the enrichment of the intragroup medium in the NGC~2300 group
indicates that stripping may be important in this system, but 
the density of the IGM is much too tenuous to effectively strip
the gas from the galaxy. However, we propose that gravitational interactions
in the group environment may enhance stripping. During a gravitational 
encounter the disk of the spiral galaxy may be warped, making ram pressure
stripping more efficient than in a quiescent disk. 

\end{abstract}

\section{Introduction}

The discovery of hot gas in clusters of galaxies presented the possibility 
that interactions between a galaxy and its environment could influence
its evolution. A galaxy's motion through the cluster gas may play a role in the 
segregation of elliptical and S0 galaxies from gas rich spirals by
stripping of the interstellar medium. Evidence that stripping does 
occur in clusters is strongest in the case of Virgo, where the spiral galaxies 
near the cluster center show a pronounced HI deficit associated with truncation 
and asymmetry of the H I disks (Cayatte et al. 1990) 
which is attributed to the hot cluster
medium stripping HI gas from the galaxies. An extreme case may be
found in NGC 5291 on the outskirts of the IC 4329 cluster, where a
large H I mass is found outside the optical galaxy and systematically
offset outwards from the cluster core, with intense star formation
seen along the ``upstream" sides of H I clumps (Longmore et al. 1979, 
Malphrus et al. 1995).
There is evidence that this process also occurs for the hot gas seen 
in elliptical galaxies. The X-ray plume from M86 is thought to be due 
to stripping as the elliptical plunges into the 
hot cluster gas (Forman Jones \& Tucker 1985; Irwin \& Sarizan 1996). 
The discovery of diffuse gas in poor groups (e.g. Mulchaey et al. 1993; Ponman 
\& Bertram 1993) raised 
the possibility that gas stripping could also occur in environments much
less dense than that of clusters. Since the group environment is less dense 
that the cluster environment and the velocity dispersions of groups are
also less, the effects of stripping should be more subtle than those
seen in clusters. Nonetheless, optical and radio work has suggested that
some spirals in rather sparse environments show evidence of interaction
with a surrounding diffuse medium. Confirmation of this process would be
important in suggesting a new arena for environmental influences
outside of rich clusters. 

We report here new multi-wavelength observations
of two of the strongest candidates for such interactions, NGC 1961 
(Arp 184) and NGC 2276 (Arp 25 and 114),
analyzed with the aim of distinguishing the gravitational effects of
surrounding galaxies from hydrodynamic effects of surrounding gas.
We trace the young and old stellar components as well as ionized,
neutral, and X-ray gas, to distinguish the response to tidal perturbations,
which is to first order similar for stars and gas, from the effects
of a gaseous disturbance, inconsequential for old stars but manifest in
low-density gas tracers.

These galaxies were selected as showing the most likely signatures of
sweeping by a hot external medium based on radio and optical evidence.
Radio studies of bright spiral galaxies (Condon 1983; Condon et al. 1990)
identified a population of spiral galaxies with extraordinarily strong disk
emission at centimeter wavelengths. Optical work shows that this is 
invariably linked to strong disk star formation, and often to interactions.
The link to galaxy interactions is hardly unusual given the abundant evidence
that tidal forces can induce star formation, but some of the most striking
examples have no plausible nearby companion. In two of the best studied 
spirals of this kind (NGC~1961 and NGC~2276), there is no 
obviously interacting nearby companion 
and yet a strong asymmetry appears in the optical and radio maps, suggesting 
that these galaxies may be interacting with the intragroup medium rather 
than another galaxy. However, the galaxy environments of such groups have
made the interpretation ambiguous in each case; there is usually a
bright galaxy close enough in projection to potentially act as a
tidal perturber, perhaps several crossing times ago so that the
interaction is not immediately apparent.

NGC 1961 is also notable for its exceptionally large linear size among
spirals (Rubin, Roberts \& Ford 1979 ; Romanishin 1983). Rubin et al. find
that the optical rotation curve implies an encircled mass above $10^{12}$ \msun,
and mention a significant role for non-circular motions. 
This large mass is confirmed by Shostak et al. (1982) who map the galaxy
using 21 cm spectral imaging. The rotation curve derived from the channel
maps is well organized and implies a mass of $\sim$ 1.5 $\times$ 10$^{12}$ \msun . 
The integrated H I map shows that the bulk of the neutral hydrogen 
emission is coincident with the optical galaxy, but the H I is also 
extended to the NW. This extension contains $\approx$10\% of the 
total H I mass. Faint spiral arms are seen in the optical image
coincident with the H I extension. Opposite the H I extension, on 
the SW side of the galaxy, the H I distribution is sharply truncated. 
This H I morphology leads Shostak et al. (1982) to interpret this 
as stripping of the H I by the hot intergalactic medium. 
They noted rough coincidence of diffuse X-ray structure in an {\it Einstein} IPC 
image with an extended H I feature, and interpreted this as evidence of the
hot stripping medium. However, ROSAT PSPC observations do not confirm the
presence of such a hot cloud (Mulchaey et al. 1996).

NGC 2276 is a very luminous Sc I spiral, in the NGC 2300 poor group.
This is the most H$\alpha$-luminous galaxy in the Kennicutt and Kent 
(1983) survey, and is remarkable not only for its disturbed morphology but
for a star formation rate high enough to have generated 4 observed
supernovae within the last 40 years, three of which were in the 
apparent ``leading edge" and close to but not necessarily directly associated
with bright H II regions
(Iskudarian \& Shakhbazian 1967, Shakhbazian 1968, Iskudarian 1968,
Treffers 1993). While the elliptical NGC 2300 and the
spiral NGC 2276 form a relatively isolated pair according to position-driven
algorithms, they are rather far apart ($\approx 150$ kpc for H$_0 = 50$
km s$^{-1}$ Mpc$^{-1}$), which at first glance makes a tidal origin for
the asymmetric disk unlikely. The galaxy gained heightened interest with the
discovery of a diffuse X-ray medium in its surroundings, the NGC 2300 
group (Mulchaey et al. 1993).

The theme of our analysis is
to ask whether we can distinguish the effects of gravitational interactions
involving neighboring galaxies from those of ram pressure driven by
a surrounding diffuse medium, and if so whether there is any strong
evidence that ram pressure is responsible for the peculiar features of
these galaxies. Both galaxies are members of poor 
groups (Maia, da Costa \& Latham 1986) so that their distorted morphologies 
might be interpreted as evidence for a recent interaction with another galaxy 
in each group.

\section{Observations}

\subsection {ROSAT X-ray Observations}

NGC~1961 and NGC~2276 were observed by the $ROSAT$ PSPC and the HRI. The 
combination of these two instruments allows the spectral and spatial 
properties of the X-ray emission for the galaxies 
in this group to be determined. The useful 
HRI exposure for NGC~1961 is almost 83 ksec and that for NGC~2276 is about 
74 ksec. A log of observations is given in Table 1. 

The $ROSAT$ PSPC observed NGC~1961 for 14.8 ksec in March of 1993. The
galaxy was 10\farcm 4 off-axis. The total exposure time for NGC~2276  is 
$\sim$23ks and is divided between three different exposures. While the 
galaxy is within the central 20\arcmin$\,$ ring of the PSPC, no useful 
spatial information was derived from these observations. However, the PSPC 
observations of NGC~1961 and NGC~2276 allow us to fit a spectral model to 
the X-ray emission from these galaxies and thus accurately determine the X-ray
flux from these galaxies. 

\subsubsection{PSPC Spectral Results}

To determine the flux from an object detected with the $ROSAT$ HRI some 
information about the spectral characteristics of the object must be known. 
The HRI provides only very crude ($\sim$ two channel) spectral information. 
However, with the addition of PSPC data a spectral model can be determined 
which can then be used to derive an accurate conversion from HRI counts 
to flux. 

The spectra for NGC~1961 and NGC~2276 were extracted from each field with the 
background determined locally. The galaxy spectrum was extracted using a 
3\arcmin$\,$ circle centered on the galaxy position. The background was taken
from an annulus with an inner radius of 3\arcmin$\,$ and an outer 
radius of 6\arcmin$\,$. 
NGC~1961 has a total of 215 net counts (0.146 counts s$^{-1}$ 0.5 - 2.0 keV)
in a 3\arcmin$\,$ circle centered on the galaxy. The spectrum is rebinned so 
that each channel has a minimum of 25 counts and the poorly calibrated 
channels (Snowden 1994) are dropped from the spectral fit. This results in 
a spectrum with 8 channels. We fit the resulting spectrum with a Raymond-Smith
plasma model with the Galactic absorption fixed at 
9.13$\times$10$^{20}$atoms cm$^{-2}$ (Starke et al. 1992).
The best fit temperature is 0.67$^{+0.19}_{-0.24}$ keV, where the errors are 
the 90\% confidence levels for a single interesting parameter. The abundance 
is constrained to be $<$3.4 solar at the 90\% confidence level. The best fit 
has $\chi^2=$2.38 for 5 degrees of freedom.  The flux from this galaxy is 
1.79$\times$10$^{-13}$ erg s$^{-1}$ cm$^{-2}$. At the assumed distance for this
galaxy (79.7 Mpc) the soft X-ray luminosity is 1.36$\times$10$^{41}$ 
erg s$^{-1}$.

NGC~2276 is in three PSPC fields and at a different position in the PSPC in each
exposure. We extracted the spectrum from each of the fields as described above 
and fit the resulting spectra simultaneously with XSPEC. 
For the initial observation of NGC~2276 (rp900161) rebinning the spectrum 
results in 5 useful channels. For the two remaining observations this 
rebinning results in 9 channels between 0.4 and 2.0 keV with sufficient counts 
to justify spectral analysis. 
The three fields yield a total of 371 counts for NGC~2276 
which corresponds to a count rate of 0.021 counts s$^{-1}$. 
We attempted to fit a Raymond-Smith plasma with Galactic absorption to the 
spectrum but the fit yields unphysical temperatures, kT $>$5 keV and 
N$_{\rm h}$ $\sim$ 0. Since the Raymond-Smith model gave unacceptable results 
we fit a simple power law model with Galactic absorption to the data. The photon
index of the best fit power law ($\chi_\nu^2$ = 0.47 for 17 dof) is -1.42. 
The X-ray flux in the 0.4 -- 2.0 keV band is 
7.26$\times$10$^{-12}$ erg cm$^{-2}$ s$^{-1}$, which for the assumed distance 
to this group of 45.7 Mpc (H$_0$ = 50 km s$^{-1}$ Mpc$^{-1}$) yields a 
luminosity of 1.82$\times$10$^{42}$ erg s$^{-1}$. The Galactic column has been 
fixed at 3.2$\times$10$^{20}$ cm$^{-2}$, the value given by Stark et al. (1992).

\subsubsection{HRI Spatial Results}

To determine any X-ray structure in these two galaxies the $ROSAT$
HRI data must be corrected for known spatial irregularities. This is
accomplished by flat fielding the images using the 
procedures outlined in Snowden et al.  (1994). The exposure maps are generated using
information of the aspect and wobble of the satellite along with a map of 
the known detector irregularities. Using this information a map can be 
generated which accurately
reproduces the spatial variations for each observation. The data and 
resulting exposure maps are binned in 5\arcsec$\,$ pixels. The data are then 
divided by the exposure map to generate the flat fielded image, which is
used in the spatial analysis below.

\subsection{NGC~1961}
The flat-fielded HRI data, with 5\arcsec$\,$ pixels and smoothed with a
Gaussian ($\sigma$=10\arcsec$\,$) , for NGC~1961 are shown overlaying a digitized 
Sky Survey (blue-light) image in figure 1. The centroid of the X-ray emission is 
at 5$^{\rm h}$42$^{\rm m}$04\fs 3 +69\arcdeg 
22\arcmin 46\farcs 3, in excellent agreement with the optical position of the 
galaxy given in the RC3. The counts for this source are extracted 
using a circle with a radius of 6\farcm 5 centered at the peak of the X-ray 
emission. The net number of HRI counts in the galaxy is 447.7$\pm$60.43 counts,
which corresponds to a count rate of 5.46$\times$10$^{-3} \pm$ 0.74 
$\times$10$^{-3}$ counts s$^{-1}$. Analysis of the X-ray profile of NGC~1961 
shows that the X-ray emission is extended. Figure 2 shows the profile of 
NGC~1961 and the profile of a point source in the same field.
The point source is only $\sim$2\farcm 5 from the position of NGC~1961 and has 
been scaled to match the counts in the inner 0\farcm 2. An excess of counts 
can been seen between 0\farcm 25 and 0\farcm 8 from the center,
and is 272 counts (3.32$\times$10$^{-3}$ 
counts s$^{-1}$). Using the flux determined from the PSPC observation of 
this galaxy  we determine the conversion factor between counts s$^{-1}$
and flux to be 0.305$\times$10$^{11}$ counts cm$^2$ erg$^{-1}$. Assuming 
that this also holds true for the diffuse component alone, we obtain a flux of 
1.09$\times$10$^{-13}$ erg s$^{-1}$ cm$^{-2}$ which at the assumed distance 
of the galaxy corresponds to an X-ray luminosity of 
8.31$\times$10$^{40}$ erg s$^{-1}$.

\subsection{NGC~2276}

The HRI data for NGC~2276 have been re-sampled into 5\arcsec$\,$ bins and then
smoothed with a Gaussian with $\sigma$=10\arcsec . Figure 3 shows the B band 
image of NGC~2276 with the  X-ray contours from the smoothed image overlaying 
the optical image. Two prominent regions of X-ray emission can be seen; the 
strongest is to the northwest of the nucleus and is a 12 $\sigma$ peak. The 
nucleus of the galaxy is also an X-ray source and is at 7$^{\rm h}$
27$^{\rm m}$21$^{\rm s}$ +85$\arcdeg$ 45$\arcmin$ 13.9$\arcsec$. 
X-ray emission from the disk of the galaxy can been 
seen around the nuclear region and along the western edge of the disk. 

The total counts from the galaxy were determined using a circle with radius 
90\arcsec$\,$ centered on the nuclear X-ray emission from the galaxy. The background 
is determined using an annular ring with an inner radius of 90\arcsec$\,$ and an
outer radius of 180\arcsec$\,$ from the center of the galaxy and all point 
sources were removed from this annulus. 
The background subtracted  HRI count rate for this galaxy is 
1.33$\times$10$^{-2} \pm$ 0.12 $\times$10$^{-2}$ counts s$^{-1}$. Using the 
spectral fit from above we can then derive the conversion between HRI counts 
and flux as 0.203$\times$10$^{11}$ counts cm$^2$ erg$^{-1}$.
Using an elliptical aperature (22\arcsec$\,$ x 37\arcsec$\,$)
centered on the peak of the emission from the 
northwest quadrant of the galaxy, we determine the count rate is 
5.58$\times$10$^{-3}\pm$ 0.73$\times$10$^{-3}$ counts s$^{-1}$,
which corresponds to an X-ray luminosity 
of 7.26$\times$10$^{41}$ erg s$^{-1}$. The count rate from the nuclear region is
2.58$\times$10$^{-3} \pm$ 0.32 $\times$10$^{-3}$ counts s$^{-1}$ which 
corresponds to an X-ray luminosity of 3.2$\times$10$^{41}$ erg s$^{-1}$ in 
a 22\arcsec$\,$circular aperature. 
This implies that the X-ray luminosity from the disk of this galaxy is at least 
7.9 $\times$10$^{41}$ erg s$^{-1}$.

\section{Optical Data}

Images in a variety of passbands and spectral data were acquired for both 
galaxies to trace both young and old stellar components. Narrow-band images, 
with filters(of typical bandwidth 60 \AA\ )
isolating H$\beta$, [O III] $\lambda 5007$, H$\alpha$+[N II] $\lambda6583$,
and [S II] $\lambda \lambda 6717+6731$ at each galaxy's redshift, along with 
adjacent continuum bands at 5125 and 6694 \AA\  for continuum subtraction,
were obtained at the KPNO 2.1m telescope with the video-camera system 
in October 1983.  The field of view was 140 arcseconds, well matched
to the size of NGC 2276 and requiring two pointings for NGC 1961 (so
that only the central part was observed in the H$\beta$, [O III], and [S II]
passbands). Additional broad-band images were obtained with CCDs for NGC 2276
(in B at the KPNO 2.1m and BVRI at the Lowell 1.1m) and NGC 1961
(BRI and narrowband for H$\alpha$). The Video Camera data used a combination 
of internal quartz and night-sky flat fields, with geometric distortions 
produced by the image-tube chain corrected after flat-field division. Since 
this corrects to constant surface-brightness (rather than point-source) 
response, the H II region fluxes required a modest correction (reaching 25\% 
only at the extreme corners) to recover accurate integrated fluxes. The 
correction (which amounts to the Jacobean of the local coordinate 
transformation) was derived from observations of photometric standard stars 
in NGC 2419. The 
device was quite stable for these observations, 
with none of the Moire patterns which sometimes plagued such data. Thus, the 
fluxes should be reliable at the 15\% level for all but the faintest H II 
regions; issues of blending and size definition are more important error 
sources than known instrumental effects.

\subsection{NGC 1961}

The optical morphology of NGC 1961 is somewhat confused by inclination
issues. The outer regions alone would indicate a substantially inclined
spiral pattern, seen perhaps 45$^\circ$ from the plane. However, the
inner isophotes are almost circular in regions devoid of obvious
extinction; the usually flattened bulge geometry of spirals would
suggest that this area is seen nearly face-on. The dust lanes cut
this bulge strongly, obviously at a significant angle to the sky plane.
One might view the system as being strongly warped, with the bulge and
inner spiral features north of the nucleus seen close to the plane of the
sky and the outer arms plus inner dust features seen about 40$^\circ$
from edge-on. The very disturbed nature of the disk makes normal
morphological classification (as well as inclination estimates)
more guesswork than we might like. The apparent pitch angles of
the arms to the north and east are inconsistent with the dust geometry
just south of the nucleus, and the more open arms to the west, for
any simple coplanar geometry.

     The disk must be far from the plane of the sky in order to give the
very large rotation velocities measured by Rubin et al. (1979), where 
the {\it observed} velocities before any inclination correction are among
the largest ever seen in a spiral disk. However, beyond this, analysis
of the geometry is limited; in the words of Rubin et al.,
``It is difficult to know just how much symmetry we can force on NGC 1961".

For NGC 1961, aperture spectra of the nucleus and several of the brightest
complexes of H II regions were obtained using the KPNO 2.1m and 
Intensified Image-Dissector Scanner (IIDS) with 6.1\arcsec$\,$ circular 
apertures and the Mount Lemmon 1.5m telescope and similar scanner 
(4.7\arcsec$\,$ apertures), in late 1983.
The line properties from the aperture spectrophotometry of both galaxies
are listed in Table 2. Numerous additional regions could be measured
in at least H$\alpha$ after calibrating the narrow-band images via
the aperture measurements (typically with 5-arcsecond apertures). Properties 
of these additional regions for both NGC 1961 and NGC 2276 are given in 
Table 3. The intent was to select the ones bright enough for meaningful
line ratios, rather than a sample complete in H$\alpha$ flux. The table
includes as well equivalent widths of H$\alpha$ emission, measured in the
same 5\arcsec$\,$ apertures, as a guide to the emission-line contrast of each
association.

A more complete listing of H II regions in NGC 1961 is provided; these are
relatively better separated and physically larger (typical H$\alpha$
FWHM 4\arcsec$\,$ ~$\approx$ 1 kpc) than we see in NGC 2276, so at the distance of 
NGC 1961 it is more appropriate to speak of H II complexes rather than 
individual H II regions.

\subsection{NGC 2276}

The line images for NGC 2276 were calibrated by means of aperture 
spectrophotometry of the brightest two H II regions, obtained using 
8\arcsec$\,$ apertures with the Intensified Reticon Scanner (IRS) at what 
was at the time (January 1984) the \#1 0.9m
telescope at KPNO. Use of such a small telescope was driven by the inability
to use the 2.1m with IIDS spectrograph within $5^\circ$ of the celestial
pole. The H II regions were acquired by blind offsets from the 8th-magnitude
star SAO 001148 (only 2.4 arcminutes from NGC 2276),
since even the galaxy nucleus could not
be seen through the guiding eyepiece. Using these data to set the absolute
intensity scales
for the line images, and assuming that the [N II]/H$\alpha$ ratios in
these areas are typical,
allows measurement of the intensities of all four lines for any
desired location from the images. A lower-resolution slit spectrum,
obtained with the KPNO 4-meter telescope and Cryogenic Camera, was
obtained with an east-west slit position crossing the nucleus, to 
get spatially continuous coverage of the [O III]/H$\beta$ ratio
and measure the [N II]/H$\alpha$ ratio with higher confidence.

The broad-band images are useful in distinguishing the young stellar
population (H$\alpha$ and blue light) from the older population which
should be unaffected by purely gas-dynamical processes (dominating the
I-band light). In NGC 2276, all show the same characteristic truncation
at the western edge (Figure 3), which in itself suggests that gravitational 
effects produce the asymmetry (as discussed by Gruendl et al. 1993). The 
eastern part of the disk, with a much lower mean star-formation rate as 
measured from H$\alpha$, appears normal, with a smooth intensity profile and no
such truncation.


Figure 4 shows the distribution of H$\alpha$ emission. H$\alpha$ imaging traces
the distribution of OB stars and hence recent star formation. Figure 4 shows 
that the H$\alpha$ flux from this galaxy is very asymmetric, with the majority 
of H$\alpha$ flux originating from the western side of the galaxy. 
The truncated side of the galaxy disk is lined by HII regions while the 
eastern half of the galaxy has only a handful of H$\alpha$ knots. The H$\alpha$
flux from the nuclear region is 1.96$\times$10$^{-13}$ erg s$^{-1}$ cm$^{-2}$.
The integrated H$\alpha$ flux from the galaxy is 3.39$\times$10$^{-12}$ 
erg s$^{-1}$ cm$^{-2}$. 

Line-ratio measurements show another aspect of this asymmetry - not only 
is the distribution of H II regions different on the two sides of the
galaxy, so is their ionization. While both sides of the galaxy show the
familiar abundance-linked gradient in [O III]/H$\beta$, this gradient
is twice as steep on the western (``upstream") side of the disk. Such
a situation might be found if, for example, disk gas on this side had been swept
inward from its original position by whatever process. A similar
azimuthal structure to the ionization gradient has been reported for
M101 (Kennicutt \& Garnett 1996), at a lower level. It may be relevant that the 
optical disk of M101 is also rather asymmetric, at a level difficult to 
reconcile with the faintness of its immediate companion NGC 5474.
We are not aware of any detailed modelling of the chemical results of
gas sweeping within a disk, so only schematic considerations can be
applied.
The diffuse gas (represented observationally by H I) will be driven inward
on the ``upstream" side, perhaps with material dropping out of the
radial flow as some gas condenses to a much denser molecular form
with correspondingly smaller cross-section for pressure-driven  acceleration.
To first order, the abundance gradient will by the ratio of
initial and final galactocentric distances, whether this distance is the 
total radial flow or the motion to dropout from radial flow. 
Such an externally-driven change in the abundance gradient will 
persist for an only orbital time or so, since clouds initially at similar radii
may have different radial velocities and thus radial diffusion driven
by their different orbits will set in. This may be seen from typical
$n$-body models for  perturbed disks, for example shown well in the
high-inclination models by Howard et al. (1993).

The optical data for NGC~2276 has been re-sampled and smoothed so that
the resolution matches that of the X-ray data and then fluxes were extracted 
from regions which match the X-ray regions. Table 4 lists the flux and 
luminosity determined for the different regions discussed in \S 4.2.
These values allow us to compare the stellar populations as revealed through their
X-ray sources as well as the direct starlight and recombination radiation.

\section{Radio Data}
VLA\footnote{The National Radio Astronomy Observatory is a facility of the 
National Science Foundation operated under cooperative agreement by Associated 
Universities, Inc. }
synthesis maps at 1.4 and 5 GHz were obtained using data from the A 
\& B array. These data were combined and cleaned using APCLN to obtain 
the final images.  The final resolution is $\sim$4\arcsec . Both these
galaxies exhibit strong and small radio sources associated with luminous
H II regions, as noted by Condon (1983).

\subsection{NGC~1961}
The radio morphology of NGC~1961 at 20cm (figure 5) is quite similar to that seen 
in the B-band image. Diffuse emission can be seen around the nucleus and a ridge of
emission can be seen about 1\arcmin$\,$ to the south which corresponds to a 
spiral arm seen in the optical image. 
The 20cm flux from NGC~1961 within 60\arcsec$\,$is 150.6 mJy.
The strongest individual source in this  galaxy is the nucleus at 10.61 mJy.
Several pointlike sources can be seen to the west of the nuclear region and these
are labeled in figure 5. Knot 1 has a flux of 4.26 mJy; knot 2's flux is 2.61 mJy; 
knot 3 has a flux of 0.94 mJy. Subtracting the flux from these sources from the total
flux from the galaxy yields a flux of 132.2 mJy for the diffuse component. 

\subsection{NGC~2276}
The 20cm radio data shown in figure 6 reflects the morphology of the 
optical data discussed above. 
The southwest edge of NGC~2276 appears to end abruptly and there are at least 
five radio bright sources within 15\arcsec$\,$ of the truncated edge of the 
radio disk. To the east no sharp boundary is seen and the radio emission 
gradually is lost in the noise. The extent of the diffuse radio emission on
the eastern side of the galaxy is approximately twice that of the truncated side.
However from figure 6 it is clear that the majority of the diffuse radio emission from 
this galaxy is from the truncated side of the galaxy. 
The total radio emission from this galaxy is 282.5 mJy. 
The nuclear region is comma shaped with the
tail of the comma appearing to lead into a spiral arm. One strong point source 
can be seen to the northwest of the nucleus along the edge of the radio 
emission. The nuclear region is the 
strongest point-like source and using a 6\arcsec$\,$ aperture we find that
the emission from the nucleus is 14.41 mJy. The point source on the northwest edge
of the radio emission has a flux of 6.27 mJy. Subtracting the point source emission from
the total gives a flux in the diffuse component of 261.8 mJy. 

     The morphological relation between H$\alpha$ and radio continuum
emission can be quantified through a ``smearing relation". Pieces of
the H$\alpha$ image were convolved with various assumed streaming
lengths, locally along the arm pitch. The best match occurs for
a typical value of 8\arcsec$\,$ (2 kpc). That is, if the particles
giving rise to the synchrotron emission at centimeter wavelengths are
mostly injected by supernovae close to the locations of present
H II regions, they travel typical distances along the large-scale
magnetic fields of 2 kpc.

     Several luminous H II regions in both NGC 1961 and NGC 2276 contain
strong nonthermal radio sources, compact on 1-arcsecond scales. The most
prominent is the bright H II region 40\arcsec$\,$ W and 19\arcsec$\,$ N of the nucleus in
NGC 2276, which is brighter than the nuclear source at 20 cm. The
reason for this exceptional emission is unclear. Supernova remnants (SNR) in
dense environments might be temporarily very bright, although we did not detect
any enhanced [S II] emission from shocks in these regions. Quantitatively,
SNR show line ratios [S II] $\lambda \lambda 6717,6731$/H$\alpha$
of 0.4 and larger (D'Odorico 1978, Dopita et al. 1984). This criterion
has proven effective in identifying SNR in galaxies of the Local Group
(Long et al. 1990), although at larger distances blending with neighboring
H II regions and the diffuse ISM reduces the purity of the samples
found in this way (Blair \& Long 1997). By this criterion, a single
emission region (number 10 in our listing) stands out both in 
[S II]/H$\alpha$ and [O III]/H$\beta$ in the direction expected for
shocked gas in a SNR. This is, however, not a detected 20-cm source;
all the radio-bright regions have typical values of [S II]/H$\alpha$
in the range 0.18--0.28. If number 10 is a single SNR, it falls in the
class of extraordinarily luminous remnants so far populated only by
the remnant in NGC 4449 (Blair et al. 1983), with comparable emission-line
luminosity but considerably lower ionization (since the NGC 4449 remnant
was not detected in [S II] by Blair et al.). Any substantial 
population of SNR accounting for the radio emission is either
masked by the surrounding emission of normal H II regions
or obscured by dust (at levels somewhat lower than the NGC 4449
SNR, so that our limits are mildly interesting but not as compelling
as one might like).

  The 21-cm HI line data, obtained with the VLA in C array, also reflect
the asymmetric morphology seen in the other wavebands.  The emission
is sharply peaked along the western edge of the galaxy (Fig 7.)
The integrated HI flux (16 Jy km/s) lies within the range of
values reported for single-dish observations of NGC~2276 (Huchtmeier and
Richter 1989).

\section{Summary of Observations}

The data for NGC~1961 show that the optical broad band and line emission 
from this galaxy are fairly symmetric in the inner regions. The X-ray and
radio images are also fairly regular. The only exception to this is 
that in the radio image the southern spiral arm appears to have enhanced 
emission when compared with the northern spiral arm. The X-ray emission
is peaked on the optical and radio nucleus of the galaxy and is mostly
point-like, but an extended component can be seen. The extended X-rays 
appear to be elongated in the east-west direction, as is the optical image
(see figure 1). At the lowest contour level, a weak tail-like feature
can be seen extending to the southeast of the nuclear region, which might
indicate that some of the hot ISM from the galaxy is being stripped. However,
since this contour is only 3 $\sigma$ above the background it should be 
interpreted with caution, and it is the opposite direction of the 
neutral hydrogen tail seen by Shostak et al. (1982).

In contrast to NGC~1961, NGC~2276 appears highly asymmetric in all 
wavebands. The western, or truncated, side has enhanced X-ray, optical,
and radio emission when compared to the eastern half. The optical emission
lines of [O III] and [S II] are restricted to the sharp boundary along the 
western limb of the galaxy. The diffuse X-ray emission can be seen along 
the truncated side of the galaxy in figure 3, along with the X-ray bright
region to the northwest of the nuclear region. The X-ray contours give 
an impression of being swept back in the same direction as the optical 
and HI images. 

\section{Discussion}

The goal of these observations is to determine if stripping of the ISM is 
occurring in these two groups. 
By using a combination of X-ray, optical, and radio imaging, along with
spectroscopy, we find
that the effects of the group is at least as complex as the analogous 
process in clusters (Moore et al. 1996). Ram pressure stripping or 
gravitational interactions alone are insufficient to explain the data. 
We find evidence that stripping of the ISM $and$ 
gravitational interactions are affecting the galaxies.

\subsection{NGC~1961}

The HI distribution in NGC~1961 is strongly suggestive that stripping of the
neutral gas is occurring (Shostak et al. 1982). However, the PSPC data for 
this group do not show the presence of hot intragroup gas  and
the upper limit for X-ray emission is $<$1.5$\times 10^{41}$ erg s$^{-1}$
(Mulchaey et al. 1996). Thus, any gas which 
is present in the group to strip the HI must be too cool to be detected 
using the PSPC data or too diffuse to be separated from the X-ray 
background.

Tracers of star formation and supernova remnants show that in the inner regions
of NGC~1961 the star formation is symmetrically distributed about the 
nuclear region. 
Low resolution radio maps show that the HI (Shostak et al. 1982) and the 
20cm emission is
compressed along the northern edge and much flatter along the southern 
edge, again suggesting that stripping may be occurring in this galaxy. 
However, the high resolution 20cm map (fig 5) shows that this may be the
result of the southern spiral arm being distorted and pulled further
(at least in projection) from the nuclear region. 
The distorted spiral arms seen in the radio map (fig 5) are also evident
in the B-band image. The stellar and the gaseous component are distorted
in the same manner, which is not plausible with ram pressure stripping.

As has been found in the optical and far-infrared bands, there is
reason to expect interactions to enhance the X-ray luminosity of
galaxies. This would happen through enhancement of star formation, with
the X-ray emission driven by massive stars and supernovae, and through
producing a global hot (expanding) medium if the star-formation rate
is high enough to produce an unbound, outflowing gas. Detailed
studies of individual objects show that these processes can be seen,
but there is yet no statistical study of the overall situation.
Observations of the well-known interacting pair NGC 4038/9 with
the {\it ROSAT} PSPC (Read et al. 1995) and {\it ASCA} (Sansom et al. 1996)
show that emission is seen both from the giant H II regions and galaxy
nuclei, and from a global, approximately bipolar gas interpreted as
an outflow, as has been seen in some more powerful IR-bright or
merging systems. At this point, it seems likely but unproven that
interactions increase the X-ray luminosity of galaxies.

In this context, we note that the
{\it Einstein} IPC X-ray luminosity (0.2 - 4.0 keV) of NGC~1961 is 
2.40$\times$10$^{41}$ erg s$^{-1}$ which is not unusual for the blue 
luminosity of the galaxy (Fabbiano, Kim \& Trinchieri 1992) and thus
the X-ray luminosity does not seem strongly enhanced.
The X-ray emission from NGC~1961 is mostly point-like in the HRI data and the peak 
emission is spatially coincident with the nucleus of the galaxy. This implies that 
most of the X-rays are from a region less than $\sim$5.8 kpc across. 
Using the total luminosity of the galaxy 
and the luminosity of the extended component we determine the point source has 
a luminosity of 7.25$\times$10$^{40}$erg s$^{-1}$. 

Since interactions are also known to enhance star formation and 
Condon, Frayer \& Broderick (1991) classify this galaxy as a starburst based on 
the ratio of infrared to radio flux, we compute the star formation rate.
The star formation rate for this galaxy can be estimated from its far-infrared
luminosity. The total far-infrared luminosity from 43 to 123 $\mu$m is given by
L$_{\rm IR}\approx$6$\times$10$^5$D$^2$(2.58f$_{60\mu m}$+f$_{100\mu m}$)
(Lonsdale et al. 1985;Thronson \& Telesco 1986) with D being the distance to the 
galaxy in Mpc. Using the measured IR 60$_{\mu m}$ flux of 6.60 mJy and the 100$_{\mu m}$ 
flux of 22.07 mJy,
we find L$_{\rm IR}$ = 1.49$\times$10$^{11}$M$_{\sun}$. From this we find 
the SFR = 31 M$_{\sun}$ yr$^{-1}$ for the high mass stars alone (Thronson \& 
Telesco 1986), which translates to about 150 \msun/yr$^{-1}$ for all
stars assuming a Salpeter IMF. While NGC 1961 is a very luminous galaxy (
listed as the
most luminous galaxy of any type in the Revised Shapley-Ames
Catalog, with $M_B=-23.7$), so that an extensive measure of SFR
such as this would be unusually high even without a burst, such
a high SFR requires a burst even for a galaxy this large and bright.
To get an intensive measure of recent SFR behavior, we derived an
integrated H$\alpha$+[N II] strength via integration in the narrow-band
images, which yields an equivalenth width of the sum of these
lines of 35.5 \AA\  , and flux of $6.3 \times 10^{-12}$ erg cm$^{-2}$ s
in H$\alpha$ alone, using the spectroscopic value for [N II]/H$\alpha$,
giving a luminosity of L(H$\alpha$=$7 \times 10^{42}$ erg s$^{-1}$
including a somewhat uncertain correction for foreground extinction.
While the H$\alpha$ luminosity corresponds to a large SFR 
(62 \msun/yr$^{-1}$ using a Salpeter IMF),
as implied by the total FIR strength, the
equivalent width corresponds to a rather mild or very protracted burst
(following, for example, Kennicutt et al. 1987).
This may thus be a starburst galaxy, though the overall SFR would
be high in any case since this is such a luminous (presumably massive)
galaxy. The high SFR might be due to an interaction, especially since the
morphology suggests disturbances in the outer spiral pattern.
While the star formation is quite widespread, this cannot argue
strongly either for or against an enhancement from interactions,
since the distribution of H II regions in clearly interacting
spirals span a wide range in radial and azimuthal distributions
(Keel et al. 1993).

The similarity of the distortions seen in the gas and stars cannot be explained 
using ram-pressure stripping. The coincidence of the gaseous and 
stellar components argues that ram-pressure stripping plays little if
any role in affecting the morphology of this galaxy in the inner regions. 
However, the HI distribution is indicative of stripping, thus it may be 
that both processes are occurring. 

If a gravitational encounter has distorted the disk of the galaxy, 
then the gas distribution can become complex forming tails and ring
structures (Moore et al. 1996). 
Even if the stars and gas are initially bent out of the plane of the galaxy 
together, this can expose a larger area of the gas disk to the intragroup 
medium and this would make ram pressure stripping much more likely. 
Also H I disks are often substantially more extended than stellar disks,
so if the group does contain
gas, it would take much less gas to strip the HI gas in the outer disturbed
portion of the disk than the gas farther in, where the starlight 
is easily detected. Thus, the group gas might remain diffuse enough to 
remain undetected in the $ROSAT$ PSPC data and still generate the 
swept-back appearance of the HI data. 
So given the plethora of data on this galaxy it seems that the most likely 
scenario is that NGC~1961 has undergone a gravitational encounter, which 
has distorted the stellar and gaseous components of the disk, disturbed
the spiral arms, slightly enhanced the star formation rate, and allowed 
ram-pressure stripping to remove the outer part of the original H I disk.

\subsection{NGC~2276}

This spiral galaxy was initially selected for study because of its
unusual optical morphology. 
The optical continuum and emission line morphology of this galaxy is 
very asymmetric, with the bulk of the line emission from the 
western side of the galaxy. This is also reflected in the radio 
maps of this galaxy where the brightest 20cm emission is confined 
to the western half of the galaxy. 

The X-ray morphology of NGC~2276 follows the optical and radio morphology
of the galaxy. In contrast to NGC~1961, the nuclear source in NGC~2276 is
not the strongest source; the disk emission along the western edge of this 
galaxy is just over twice the luminosity of the nuclear region. The total
log(L$_{\rm x}$/L$_{opt}$) is -1.82 while the average L$_{\rm x}$/L$_{opt}$
for late type spirals (T=4-10) is -6.90$\pm$0.27 (Fabbiano, Gioia \& 
Trinchieri 1988). The log(L$_{\rm x}$/L$_{FIR}$) is -2.36,
close to -3 as expected for normal galaxies or LINERS (Green, Anderson 
\& Ward 1992) but not typical of starburst galaxies 
which have log(L$_{\rm x}$/L$_{FIR}$) $\approx -4$ (Heckman, Armus
\& Miley 1990). 

The star formation rate can be estimated from the H$\alpha$ luminosity,
which for  NGC~2276 is 3.39$\times$10$^{41}$ erg s$^{-1}$.
Using SFR=7.07$\times$10$^{-42}$ $\eta^{-1}$L$_{\rm H\alpha}$ M$_{\sun}$ yr$^{-1}$ 
(Hunter et al. 1986) and $\eta$= 0.5,
we estimate a star formation rate of $\sim$ 5 M$_{\sun}$ yr$^{-1}$. Using the measured IR 
60$_{\mu m}$ flux of 11.97 mJy and the 100$_{\mu m}$ flux of 28.96 mJy, we
find that the SFR=15M$_{\sun}$ yr$^{-1}$.
This is in reasonable agreement with the SFR derived from the ${\rm H\alpha}$
flux given that these estimates are likely to be accurate only to within
a factor of two. As in the case of NGC 1961, it is useful to consider
an intensive quantity such as H$\alpha$ equivalent width to assess the
history of the SFR. The integrated spectral data presented by 
Kennicutt (1992) give an H$\alpha$ equivalent width of 59 \AA\  ,
to be compared with the 32 \AA\  from aperture photometry by
Kennicutt et al. (1987). This falls well above the range populated
by smoothly declining SFR models (as shown in Kennicutt et al. 1987),
putting this in the global-burst category.

The unusual morphology of this galaxy has been attributed to either ram pressure
stripping or from a gravitational interaction with NGC~2300, an elliptical with
signs of a recent merger (Forbes \& Thomson 1992, who find evidence that
it hosts a cooling flow). We have examined this
object further, via analysis of a Lowell $I$-band image, to confirm the
reported shell-like morphology. By subtracting the best overall $r^{1/4}$
model, we confirm the tidal extension and western ``shell" reported by Forbes 
\& Thomson, and detect as well a nearly symmetric ``bowtie" structure
within about 15\arcsec$\,$ of the nucleus (the edges of which appear in their residual 
image as $4 \theta$ residuals). All of these features are evidence for various 
levels of gravitational interaction.

Davis et al. (1996) and Mulchaey et al. (1993) show that the density of the 
intragroup medium itself is too small
to significantly perturb the ISM of the galaxy. Gruendl et al. (1993), using 
Fabry-Perot observations of the H$\alpha$ velocity field, conclude that a tidal
interaction with NGC~2300 is the most likely explanation for the truncated 
stellar disk and H$\alpha$ velocity field. Given the large projected separation
between the galaxies, about 150 kpc, if the transverse relative velocity
is in the range expected from velocity dispersions in small groups $< 400$
km s$^{-1}$, the encounter would have been nearly parabolic and taken place
more than $4 \times 10^8$ years ago. This is a long time for immediate
interaction effects to play out, suggesting that perhaps internal
disk modes excited by the interaction continue to enhance the star
formation even well after the perturber has departed.

From these lines of evidence, it appears that the morphology of 
NGC~2276 is primarily being affected by a gravitational encounter 
which has disturbed the stellar and gaseous component, enhanced the
star formation rate along the truncated side of the galaxy, and again
pulled the HI out of the plane of the galaxy so that stripping is much 
more likely. 
The vigorous star formation along the truncated side of the galaxy may
be enhanced by ram pressure but it cannot be the dominate force shaping
the morphology. The interaction
might be important in making some of the disk H I much more vulnerable
to stripping than it would have been originally, so that the combination
of two effects is not complete coincidence.

\section{Conclusions}

We have used the $ROSAT$ HRI to study the X-ray emission from the 
disk of two spiral galaxies in small groups, NGC~1961 and NGC~2276. To the 
best of our knowledge these are the most distant spiral galaxies to 
have their X-ray disk emission resolved. The X-ray structure of NGC~1961 
consists of a nuclear point-like source and diffuse emission from 
the disk. The nuclear source has an X-ray  luminosity consistent with that
of a low level AGN. The X-ray luminosity of the disk is 
8.31$\times$10$^{40}$ erg s$^{-1}$ and is consistent with that seen from 
other normal spiral galaxies. 
 
The disk emission from NGC~2276 is clumpy and is strongly correlated
with the bright star forming regions seen on the western side of the 
galaxy. The X-ray luminosity of the disk component is 
7.9 $\times$10$^{41}$ erg s$^{-1}$, again not unusual for spiral 
galaxies of this blue magnitude. The nuclear region is also detected
with a luminosity of 3.2$\times$10$^{41}$ erg s$^{-1}$ indicating that
this galaxy may also have an active nucleus. 

Tidal interactions have been shown to enhance the star formation in galaxies 
on both nuclear and global scales (Keel et al. 1985, Bushouse 1987,
Kennicutt et al. 1987; also see references in Keel 1991)
to an extent which is consistent with the H$\alpha$ luminosity and star 
formation rates derived above. So from the evidence we have presented above it 
appears that the strong star formation and distorted morphologies of these 
two spiral galaxies is most likely due to a tidal interaction with a companion 
galaxy and not due to ram pressure effects. 
However, the swept back appearance of the HI gas indicates that 
ram pressure stripping may be occurring. 

NGC~1961 also shows evidence for ram pressure stripping with an  HI extension
to the northeast (Shostak et al. 1982). The distorted spiral
arms indicate that this galaxy has also had a gravitational encounter; however,
no likely companion is apparent. 
Since, the X-ray observations do not reveal the presence of a dense 
intragroup medium that could strip the quiescent HI gas we believe that this
may be another example where stripping is made more efficient after a gravitational
encounter.

We began this study in the hope that selecting galaxies with the
extremes of asymmetry and star formation expected for interaction with
a surrounding diffuse medium would furnish the best evidence of
gas stripping in a relatively simple environment. As it happened,
even in these carefully picked instances, evidence for stripping is
subtle, and the group environments are rich enough that gravitational
interactions with other members still dominate the morphology and
star-forming properties of these spirals. 
The case of NGC 2276 suggests that seeing a combination of these two effects may not be
entirely fortuitous -- the tidal disturbance may render an H I disk
more vulnerable to external hydrodynamical forces than it would otherwise
be. The difficulty of separating these effects in nearby galaxies in
rather simple environments may serve as a cautionary tale for untangling
such effects in clusters and their role in galaxy evolution.

\acknowledgments
This research  made use of the HEASARC, NED, and SkyView databases. We 
acknowledge support from NASA ROSAT grant NAG5-2703. Some of the VLA data
were obtained in collaboration with Jim Condon, who also reduced the
radio-continuum data.

\newpage 
\centerline{Figure Captions}
 
\noindent{Fig. 1--The HRI data (contours) is overlaying the optical image of 
NGC~1961 from the digitized sky survey. The X-ray data have been smoothed with a 
Gaussian with $\sigma$=10\arcsec$\,$ and the contour levels are ( 6.66,
7.77, 8.88, 9.98, 11.09, 12.20, 13.31, 14.42, 15.53, 16.64)
$\times$10$^{-2}$ counts s$^{-1}$ arcmin$^{-2}$. }

\noindent{Fig. 2--The azimuthally averaged X-ray profile of NGC~1961 (triangles) along
with the azimuthally average profile of a stellar object from the same image (circles). 
NGC~1961 has excess emission from 0\farcm 25 to $\sim$0\farcm 7. 
}

\noindent{Fig. 3--The HRI data (contours) is overlaying the optical image of 
NGC~2276 from the 2-meter telescope at Kitt Peak. The X-ray data have been smoothed with a 
Gaussian with $\sigma$=10\arcsec$\,$ and the contour levels are (1.08, 1.21,
1.34, 1.46, 1.59, 1.72, 1.85, 1.98, 2.10, 2.23, 2.36)
$\times$10$^{-2}$ counts s$^{-1}$ arcmin$^{-2}$. }

\noindent{Fig. 4--The H$\alpha$ image of NGC~2276. North is up and east is to the 
left. The FOV is 2\arcmin$\,$ across. 
}

\noindent{Fig. 5--The 20 cm radio image of NGC~1961.
}

\noindent{Fig. 6--The 20 radio image of NGC~2276. Note the sharp edge of the radio 
disk to the west of the nucleus.
}

\noindent{Fig. 7.--The 21-cm HI line image of NGC 2276.  
The maps have been smoothed to a resolution of 30\arcsec$\,$ x 18\arcsec$\,$ (twice
the original resolution) with long axis at position angle -10 degrees.
Contours are the integrated
HI emission with levels at 12, 18, 24, 30, ... 90 mJy/beam.
The greyscale represents the velocity field.

\end{document}